\RequirePackage{fix-cm}
\documentclass[smallextended]{svjour3}	
\smartqed 
\usepackage[dvipdfmx]{graphicx}
\usepackage{subfigure}
\usepackage{pdfpages}
\begin{document}

\title{Location-Sector Analysis of International Profit Shifting on a Multilayer Ownership-Tax Network \thanks{The present study was supported by the Ministry of Education, Science, Sports, and Culture, Grants-in-Aid for Scientific Research (B), Grant No. 17KT0034 (2017-2019) and Exploratory Challenges on Post-K computer (Studies of Multi-level Spatiotemporal Simulation of Socioeconomic Phenomena).}
}
\author{Tembo Nakamoto \and Odile Rouhban \and Yuichi Ikeda}
\institute{Tembo Nakamoto \at
              Graduate School of Advanced Integrated Studies in Human Survivability\\
              1, Yoshida-Nakaadachi-cho, Sakyo-ku, Kyoto, 6068306, Japan\\
              Tel: +81-75-762-2002\\
              \email{nakamoto.tembo.75w@st.kyoto-u.ac.jp}
           \and
           Odile Rouhban \at
           \email{odile.rouhban@gmail.com}
\and
Yuichi Ikeda \at
              Graduate School of Advanced Integrated Studies in Human Survivability\\
\email{yuichi.ikeda.2w@kyoto-u.ac.jp}
}
\titlerunning{Location-Sector Analysis} 
\date{Received: date / Accepted: date}

\maketitle

\begin{abstract}
Currently all countries including developing countries are expected to utilize their own tax revenues and carry out their own development for solving poverty in their countries. However, developing countries cannot earn tax revenues like developed countries partly because they do not have effective countermeasures against international tax avoidance. Our analysis focuses on treaty shopping among various ways to conduct international tax avoidance because tax revenues of developing countries have been heavily damaged through treaty shopping. To analyze the location and sector of conduit firms likely to be used for treaty shopping, we constructed a multilayer ownership-tax network and proposed multilayer centrality. Because multilayer centrality can consider not only the value flowing in the ownership network but also the withholding tax rate, it is expected to grasp precisely the locations and sectors of conduit firms established for the purpose of treaty shopping. Our analysis shows that firms in the sectors of Finance \& Insurance and Wholesale \& Retail trade etc. are involved with treaty shopping. We suggest that developing countries make a clause focusing on these sectors in the tax treaties they conclude.
\end{abstract}

\section{Introduction}
\label{intro}
About 800 million people worldwide still cannot enjoy food, safe drinking water, and clean sanitation. The United Nations has established the Sustainable Development Goals (abbreviated as SDGs) with poverty as its first target. I order to provide public services required to solve poverty, such as health, education, and social security, each developing country needs funding equivalent to 4\% of their Gross Domestic Product (abbreviated as GDP) (UN 2005). So far, such funding has mostly relied on developed countries. Recently, however,  it has been emphasized that it is important for developing countries to procure such funds by themselves (G20 2010). It has been confirmed again in the implementation of SDGs (UN 2015) and it is set as SDG 17.1 to develop the taxing capabilities of developing countries. To reduce poverty, developing countries are expected to increase their tax revenue.\par
However, developing countries' ability to raise tax revenues is still lower than that of developed countries. While developed countries earn more than 30\% tax on their GDP, many developing countries do not reach as much as 20\% except some Asian and Latin American countries. In particular, for more than half of sub-Saharan African countries, it is less than 15\% (UNDP 2010). The reason why developing countries cannot earn enough tax revenue for the scale of their economies is mainly due to a lack of tax collection capacity (Gordon and Li 2009) and international tax avoidance. In particular, low-income countries in sub-Saharan Africa, Latin America, the Caribbean, and South Asia have lost much of their tax revenues due to international tax avoidance (Cobham and Jansky 2017). Compared with developed countries, developing countries are more vulnerable to international tax avoidance because many developing countries do not have effective policies to prevent international tax avoidance (OECD 2014). It is needed now to craft an effective policy toward international tax avoidance for developing countries to earn adequate tax revenues.\par
We focused on treaty shopping among various ways to avoid taxes. Treaty shopping is a scheme to avoid withholding tax. When a firm pays its dividends to a firm in another jurisdiction, withholding tax is usually imposed on the dividends. The withholding tax is sometimes reduced or exempted when the payment is conducted between jurisdictions contracted in a tax treaty. Treaty shopping occurs when a firm in a third jurisdictions and not eligible to receive the reduction or exemption tries to establish a so-called paper company in tje contracting jurisdiction in order to enjoy the reductions or exemptions. Many developing countries have reported huge losses in their tax revenue as a result of treaty shopping (OECD 2014). Our analysis deals with treaty shopping, which heavely damages developing countries by loss of their tax revenues.\par
Recently, network science has attracted attention because network theory has succeeded in highlighting emergent phenomena called "complex systems,"  which describes economic phenomena with a graph consisting of nodes and links (Barabasi 2016). However, actual economic phenomena often have to consider multiple layers, as a result, research on multilayer networks has progressed (Kivela et al. 2014). Multilayer networks consist of layers in which the nodes of one layer are connected to nodes in another.\par
There have been only a few previous studies on treaty shopping. Mintz and Weichenrieder (2008) and Weyzig (2013) investigated treaty shopping taking into account both the profits and the withholding tax rates, but their target is only specific countries (Germany and The Netherlands). Garcia-Bernardo et al. (2017) investigate treaty shopping by focusing the profits (value) in the world and Van't Riet and Lejour (2018) focuses on the withholding tax rates of 108 jurisdictions. While Van't Riet et al. (2015) examined the relationship between treaty shopping and dividends, Hong (2018) and Petkova et al. (2018) studied the effects of double tax treaties on foreign direct Investment. There are no studies to analyze treaty shopping taking both profits and withholding tax rates worldwide. In addition, unlike other methods of international tax avoidance (Acciari et al. 2015), there are no studies to investigate treaty shopping from the point of view of sectors. This is the first study to analyze treaty shopping from the perspective of sectors, while taking into account both profits (values) and the withholding tax rate worldwide.\par
We construct and analyze a multilayer network consisting of an ownership network and a withholding tax network. There are three features of our analysis. The first point is the use of micro data to record firm information all over the world. The second point is that we analyze treaty shopping by focusing on sectors for the first time. Third, the multilayer centrality we propose makes it possible to analyze treaty shopping more precisely by considering both the profit (value) and the withholding tax rate.\par
Section 2 describes the multilayer network we construct and explains multilayer centrality, which we propose for analyzing the locations and sectors of firms used for treaty shopping. Section 3 outlines the data used to build the multilayer network. Section 4 describes the results obtained by applying our multilayer centrality to the multilayer network and considers effective tax policies to prevent treaty shopping. Section 5 summarizes our study.\par

\section{Method}
\label{sec:2}
Although countermeasures against some international tax avoidance methods have already focused on specific sectors, countermeasures against treaty shopping have not focused on the sectors. We supposed that there are some sectors that are easy to exploit for treaty shopping like other tax avoidance methods and examined the locations and the sectors of conduit firms used in treaty shopping. The concept of "value" is used for our analysis because its effectiveness has been confirmed in previous research (Garcia-Bernardo et al. 2017). However, the purpose for which conduit firms are established is not limited to the use of treaty shopping. To limit our analysis to conduit firms for treaty shopping, we also considered withholding tax rates imposed on dividends. One of the main motivations to do treaty shopping is to recommend withholding tax rates. We propose multilayer centrality as means to analyze the conduit firms used in treaty shopping while taking into account both the value intensity and the withholding tax rates.

\subsection{Multilayer ownership-tax network}
\begin{figure}[ht]
\centering
\includegraphics[width=8cm]{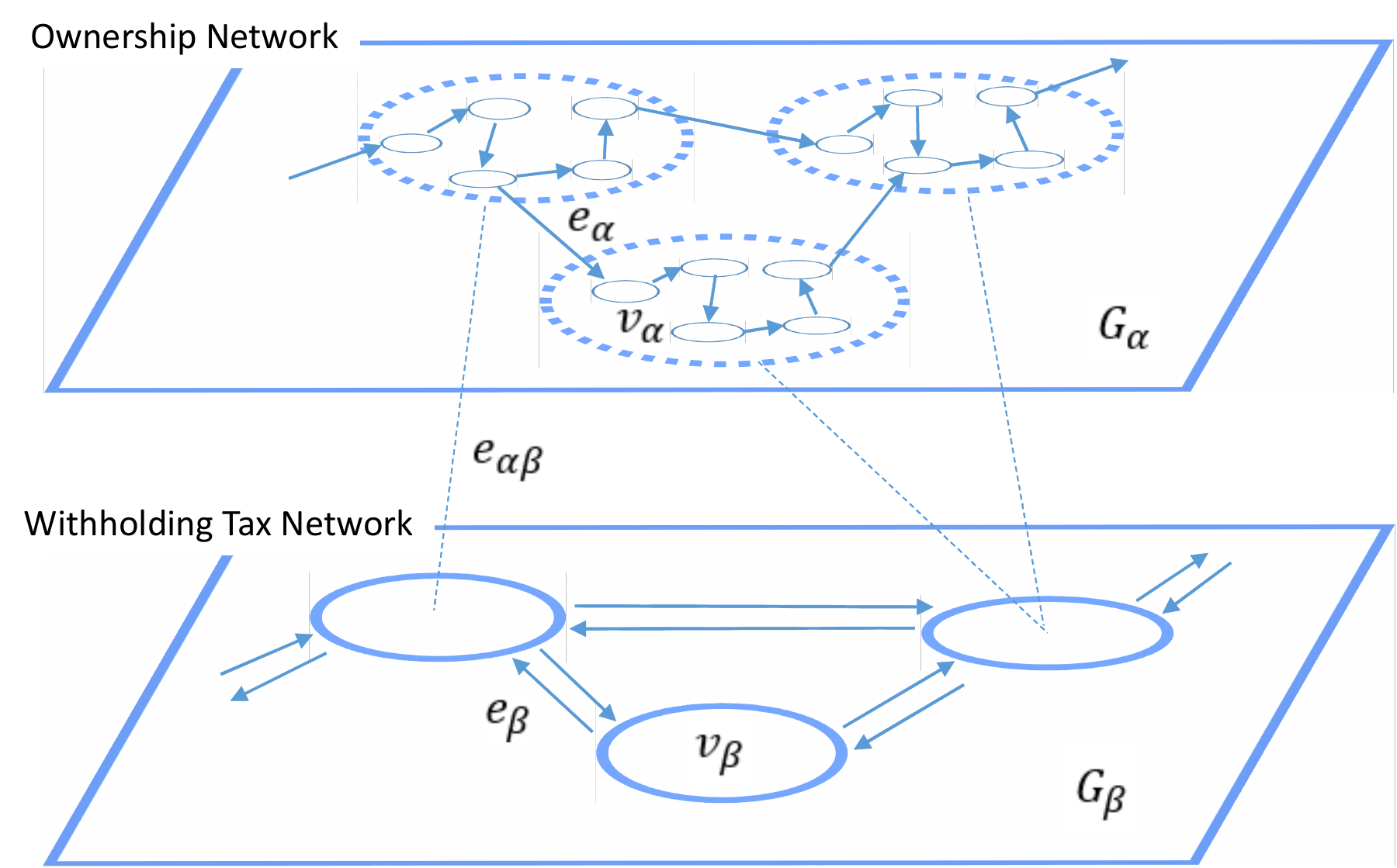}
\caption{Image of the multilayer ownership-tax network.}
\label{fig:1}       
\end{figure}

The multilayer ownership-tax network is defined as $M = (g, l)$, where $ g = \lbrace G_\alpha, G_\beta \rbrace$ is a set of weighted directed graphs (called layers of $M$) and $l \in L_{\alpha\beta}$ is a set of interlayer connections. Figure \ref{fig:1} shows an image of the network $M$. The first layer of $M$, defined as $G_\alpha = (N_\alpha, L_\alpha, W_\alpha,  P_\alpha)$, is the ownership network, where $N_\alpha$ is a set of nodes (firms), $L_\alpha$ is a set of links (shareholding relationships) between pairs of nodes, $W_\alpha$ is the link value function (shareholding ratios), and $P_\alpha$ is the node value function (operating incomes). The links are directed, going from a shareholder firm to an owned firm. The second layer of $M$, defined as $G_\beta=(N_\beta , L_\beta, W_\beta)$, is the withholding tax network, where $N_\beta$ is a set of nodes (jurisdictions), $L_\beta$ is a set of links (direction to pay dividends) between pairs of nodes, and $W_\beta$ is the link value function (withholding tax rates imposed on dividends. A node (firm)  $n_\alpha$ of the layer $G_\alpha$ is connected with that (jurisdiction) $n_\beta$ of the layer $G_\beta$ according to the location of $n_\alpha$ by $L_{\alpha\beta}$.

\subsection{Value intensity}
The firms used for tax purposes are divided into "sink" and "conduit" according to their functions (Garcia-Bernardo et al. 2017). We analyze the conduit firms used for treaty shopping by focusing on value intensity. The sink firms have to be identified for analyzing the conduit firms because of the conduit firms' definition. Therefore, we first identify the locations and sectors of firms functioning as a "sink" and then analyze the locations and sectors of firms functioning as a "conduit."

\subsubsection{Value}
\label{sec:2-2-1}
To identify the sink and the conduit, we use the value flowing in the ownership network. The value is defined as follows (Vitali et al. 2012):
\begin{equation}
v_{n_{i + 1}} = p_k \prod_{i = 1}^{l - 1} w_{n_i n_{i + 1}}
\end{equation}
Here, $p_k$ is an operating income of a firm $k$ located at the end of chains in the ownership network and $w_{n_i n_{i + 1}}$ is the shareholding ratio from a shareholder firm $n_{i + 1}$ to an owned firm $n_i$. The value $v_{n (i + 1)}$ enters a firm $n_{i + 1}$ and the value $v_{n (i + 2)}$ leaves firm $n_{i + 1}$.\par
The locations and sectors of firms are expressed by jurisdiction $\times$ sector pairs. For example, The Netherlands $\times$ Finance \& Insurance indicates a set of firms located in The Netherlands whose sector is financial and insurance activities.

\subsubsection{Sink}
\begin{figure}[ht]
\centering
\includegraphics[width=8cm]{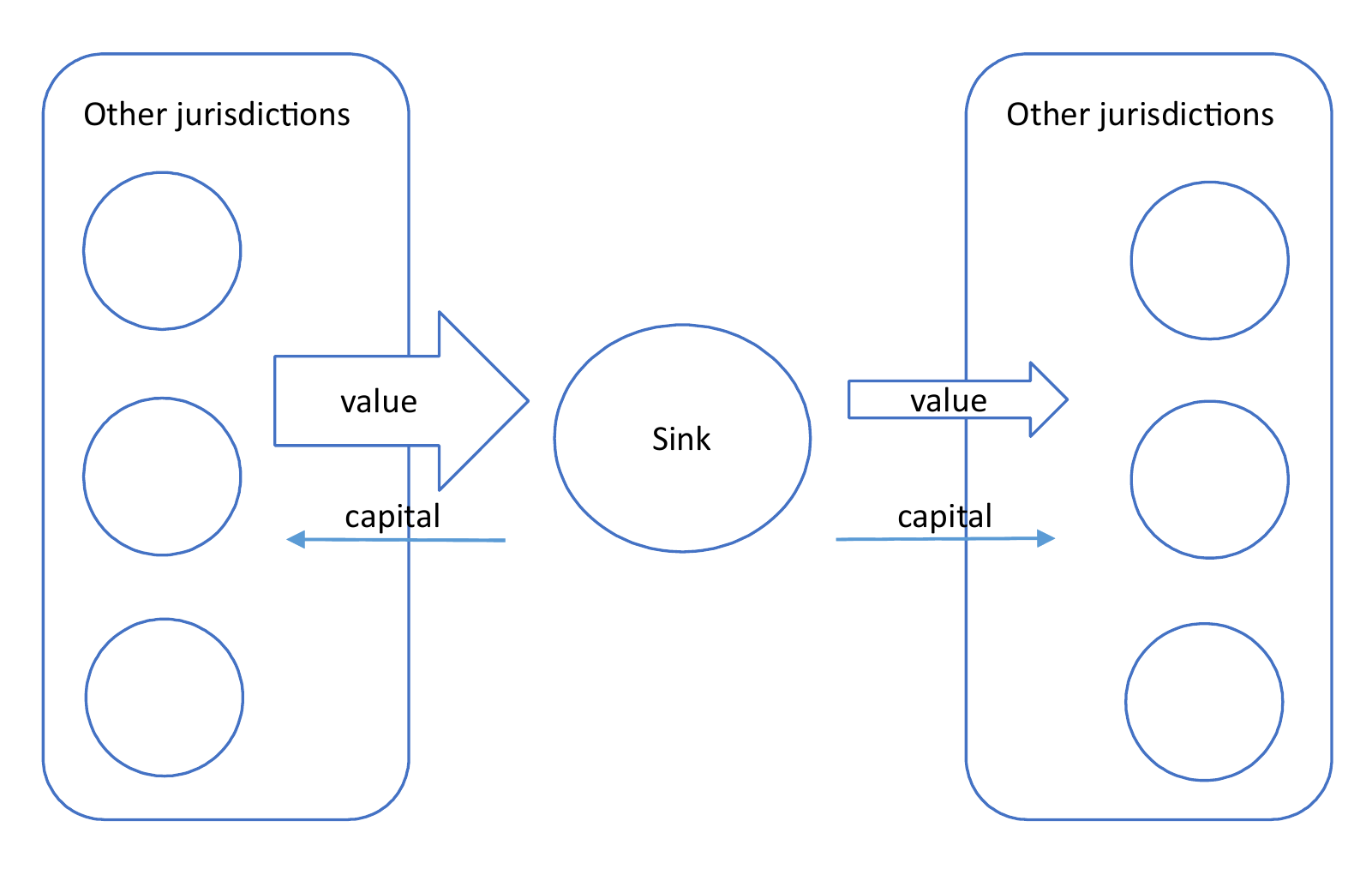}
\caption{{\bf Sink concept}: more value enters jurisdictions $\times$ sector pairs functioning as a sink and less value leaves the pair.}
\label{fig:2}       
\end{figure}

Figure \ref{fig:2} shows the concept of a "sink." We supposed that a jurisdiction $\times$ sector pair functioning as a "sink" has much more value compared with its economic scale. 

The sink centrality $S_{js}$ of jurisdiction $j$ $\times$ sector $s$ pair is defined as follows. At first, take the difference between the value entering and the value leaving firms whose location is $j$ and sector is $s$ to calculate the value of the pair. Next, this is divided by the total value flowing in the ownership network to calculate the value intensity of the pair. Finally, it is normalized by jurisdiction GDP to compare the value intensity with its economic scale:
\begin{equation}
S_{js}=\frac{\sum V_{js}^{in} - \sum V_{js}^{out}}{V^{total}} \cdot \frac{\sum_i GDP_i}{GDP_j}
\label{eq:1}
\end{equation}
Here, $\sum V_{js}^{in}$ is the sum of the values entering the firms whose location is $j$ and sector is $s$, $\sum V_{js}^{out}$ is the sum of the values leaving the firms whose location is $j$ and sector is $s$, $V^{total}$ is the total amount of the value flowing in the ownership network, and $GDP_j$ is the GDP of jurisdiction $j$. We suppose that the GDP of each jurisdiction represents its economic scale.\par

\subsubsection{Conduit}
\begin{figure}[ht]
\centering
\includegraphics[width=8cm]{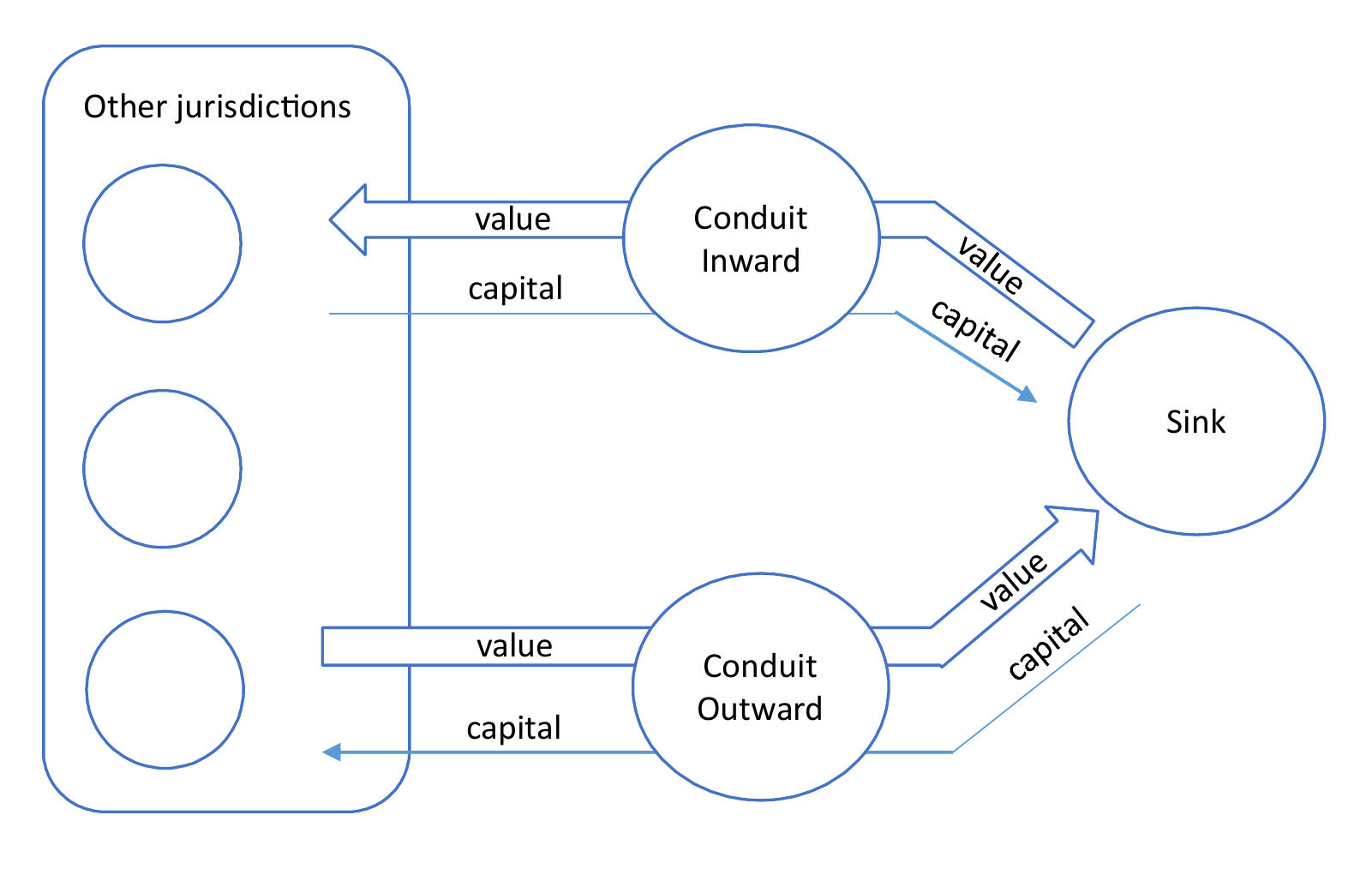}
\caption{{\bf Conduit concept}: value enters into "sink" through "conduit outward" and leaves "sink" through "conduit inward."}
\label{fig:3}       
\end{figure}

A conduit is like a tunnel through which value enters or leaves a sink. We consider that much more value entering the sink or leaving the sink passes through jurisdiction $\times$ sector pairs functioning as a conduit compared with its economic scale. Because the conduit plays a key role in treaty shopping, we analyze the location and sectors of firms functioning as conduits. For the analysis, we define a conduit centrality $c_{js}$ of jurisdiction $j$ $\times$ sector $s$ that consists of conduit outward centrality $c_{js}^{out}$ and conduit inward centrality $c_{js}^{in}$. Figure \ref{fig:3} shows the concept of a conduit. The conduit outward centrality $c_{js}^{out}$ measures the value entering the sink through firms whose location is $j$ and sector is $s$ while the conduit inward centrality $c_{js}^{out}$ measures the value leaving the sink through firms whose location is $j$ and sector is $s$. 

The conduit outward centrality $c^{out}_{js}$ of jurisdiction $j$ $\times$ sector $s$ is defined as follows. At first, it measures the values passing through firms whose location is $j$ and sector is $s$ and entering the sink. Next, it is divided by the total amount of values flowing in the ownership network $G_\alpha$ to calculate the value intensity. Lastly, it is normalized by jurisdiction GDP to compare the passing value with its economic scale.
\begin{equation}
c_{js}^{out}=\frac{V_{out}^{sink}}{V_{total}} \cdot \frac{\sum_i GDP_i}{GDP_j}
\label{eq:3}
\end{equation}
Here, $V_{out}^{sink}$ is the sum of the value passing through the firms whose location is $j$ and sector is $s$ and entering the sinks, $V_{total}$ is the total amount of the value flowing in the ownership network $G_\alpha$, and $GDP_j$ is the GDP of jurisdiction $j$ representing its economic scale.\par
The conduit outward centrality $c_{js}^{out}$ is standardized so that its average value and standard deviation is 1.0 because of the calculation of the multilayer centrality described in Sect.~\ref{sec:10}:
\begin{equation}
C_{js}^{out} = \frac{c_{js}^{out} - \overline{c_{js}^{out}}}{\sigma_{c_{js}^{out}}} + 1
\label{eq:6}
\end{equation}
Here, $\overline{c_{js}^{out}}$ is the average of all conduit outward centrality $c_{js}^{out}$ and $\sigma_{c_{js}^{out}}$ is the standard deviation of all conduit outward centrality $c_{js}^{out}$. Therefore, if the standardized conduit outward centrality $C_{js}^{out}$ of a jurisdiction $j$ $\times$ sector $s$ pare is above 1.0, then it can be said that the value to sink has passed through the jurisdiction $j$ $\times$ sector $s$ pare while much considering its economic scale compared with other pairs.\par

On the other hand, the conduit inward centrality $c^{in}_{js}$ of jurisdiction $j$ $\times$ sector $s$ is defined as follows. At first, it measures the values leaving from the sink and passing through the firms whose location is $j$ and sector is $s$. Next, it is divided by the total amount of values flowing in the ownership network $G_\alpha$. Lastly, it is normalized by jurisdiction GDP to compare the passing value with its economic scale:
\begin{equation}
c_{js}^{in}=\frac{V_{in}^{sink}}{V_{total}} \cdot \frac{\sum_i GDP_i}{GDP_j}
\label{eq:2}
\end{equation}
Here, $\sum V_g$ is the sum of the value from sink through firms whose location is jurisdiction $j$ and sector $s$, $\sum V_g$ is the total amount of the values flowing in the ownership network $G_\alpha$, and $GDP_j$ is the GDP of jurisdiction $j$ representing its economic scale.\par
The conduit inward centrality $c_{js}^{in}$ is standardized so that its average value and standard deviation is 1.0 because of the calculation of the multilayer centrality described in Sect.~\ref{sec:10}:
\begin{equation}
C_{js}^{in} = \frac{c_{js}^{in} - \overline{c_{js}^{in}}}{\sigma_{c_{js}^{in}}} + 1
\label{eq:16}
\end{equation}
Here, $\overline{c_{js}^{in}}$ is the average of all conduit inward centrality $c_{js}^{in}$ and $\sigma_{c_{js}^{in}}$ is the standard deviation of  all conduit inward centrality $c_{js}^{in}$. Therefore,  if the standardized conduit inward centrality $C_{js}^{in}$ of a jurisdiction $j$ $\times$ sector $s$ pare is above 1.0, then it can be said that the value from the sink has passed through the jurisdiction $j$ $\times$ sector $s$ pare while considering its economic scale compared to other pairs.\par

To make it easy to compare between jurisdiction $\times$ sector pairs, the Euclidean distance between the standardized conduit outward centrality $C^{out}_{js}$ and the standardized conduit inward centrality  $C^{in}_{js}$ is calculated. The distance is adjusted to 1.0 when the standardized conduit outward centrality $C^{out}_{js}$ and the standardized conduit inward centrality $C^{in}_{js}$ are both 1.0: 
\begin{equation}
C_{js} = \sqrt{(C_{js}^{in})^2+(C_{js}^{out})^2}/\sqrt{2}
\label{eq:4}
\end{equation}

\subsection{Withholding tax rate}
Conduit firms are established for a variety of tax reasons. The purpose of our analysis is to analyze the conduit firms established for treaty shopping. Therefore, we focused on the withholding tax imposed on dividends, which is one of the most important factors in conducting treaty shopping (Weyzig 2013). We regard conduit firms located in jurisdictions highly attractive to treaty shopping from the point of the withholding tax rates as the conduit firms used for treaty shopping.\par
We used load centrality $l_j$ to measure the attraction of treaty shopping quantitatively. Load centrality $l_j$ is the total amount of the packet passing through a node when all pairs of nodes send and receive a data packet between them (Goh et al. 2001; Brandes 2008). We think of a node as a jurisdiction, a data packet as a dividend, put the withholding tax rates as the weight, and calculate the load centrality $l_j$ (Nakamoto and Ikeda 2018). Therefore, the higher the load centrality $l_j$, the more likely it will be for the centrality to be used for treaty shopping. The load centrality $l_j$ of a node $j$ is calculated as follows:
\begin{equation}
l_j = \sum_{o \neq d \neq j} w_{o,d}
\label{eq:5}
\end{equation}
Here, $j$, $o$, and $d$ are nodes (jurisdictions), $w_{o,d}$ is the amount of the dividend passing through $j$ when a firm located in $o$ sends a dividend to a firm located in $d$.\par
The load centrality $l_i$ is standardized so that both its average value and standard deviation are 1.0:
\begin{equation}
L_j = \frac{l_j - \overline{l_j}}{\sigma_{l_j}} + 1
\label{eq:18}
\end{equation}
Here, $\overline{l_j}$ is the average of all load centrality $l_j$ and $\sigma_{l_j}$ is the standard deviation of  all load centrality $l_j$. Therefore, if the standardized load centrality $L_j$ of a given jurisdiction $j$ is above 1.0, then it can be said that jurisdiction $j$ has more possibility to be chosen as the location of conduit firms compared with other jurisdictions.

\subsection{Multilayer centrality}
\label{sec:10}
Multilayer centrality $M_{js}$ consists of the multilayer outward centrality $M_{js}^{out}$ and the multilayer inward centrality $M_{js}^{in}$ like the conduit centralities $c_{js}$. The multilayer outward centrality $M_{js}^{out}$ and the multilayer inward centrality $M_{js}^{in}$ are,  respectively, the weighted geometry average of the values of the standardized conduit outward centrality $C_{js}^{out}$ and the standardized conduit inward centrality $C_{js}^{in}$ in the ownership network $G_\alpha$ multiplied by the standardized load centrality $L_j$ in the withholding tax network $G_\beta$. The multilayer outward centrality $M_{js}^{out}$ and the multilayer inward centrality $M_{js}^{in}$ are defined as follows:
\begin{equation}
M_{js}^{out} = \sqrt[\alpha + \beta]{(C_{js}^{out})^{\alpha} \cdot (L_w)^{\beta}}
\label{eq:8}
\end{equation}
\begin{equation}
M_{js}^{in} = \sqrt[\alpha + \beta]{(C_{js}^{in})^{\alpha} \cdot (L_w)^{\beta}}
\label{eq:7}
\end{equation}
Here, $\alpha$ and $\beta$ determine the ratio to consider the value intensity and the withholding tax rate to analyze the conduit firms.\par
To make it easy to compare between jurisdiction $\times$ sector pairs, the Euclidean distance between the multilayer outward centrality $M^{out}_{js}$ and the multilayer inward centrality $M^{in}_{js}$ is calculated. The distance is adjusted to 1.0 when the multilayer outward centrality $M^{out}_{js}$ and the multilayer inward centrality $M^{in}_{js}$ are both 1.0: 
\begin{equation}
M_{js} = \sqrt{(M_{js}^{in})^2+(M_{js}^{out})^2}/\sqrt{2}
\label{eq:9}
\end{equation}

\section{Data}
\label{sec:3}
We used the Orbis 2015 database  (Bureau van Dijk 2015) for the ownership network $G_\alpha$. The database comprises shareholding ratio, operating income, and sector information of about more than 30 million firms across more than 20 jurisdictions. Sectors are defined following the statistical classification of economic activities in the European community (NACE Rev. 2).\par
This database is based on information that each firm reported to their local Chamber of Commerce. Because each jurisdiction requires different criteria for the Ministry of Commerce to submit financial statements, data availability varies greatly depending on jurisdictions. For example, Kosovo includes 99.1\% while Seychelles includes only 0.1\%. In particular, it does not include information on small-scale firms (Kalemi-Ozan et al. 2015) and information on firms located in jurisdictions where financial secrets are high. Therefore, our analysis has a certain bias due to data availability. The shareholding ratio, operating income, and sector information are necessary to calculate the value (see~Sect. \ref{sec:2-2-1}). We removed the nodes $v_\alpha$ (firms) that do not have such information from the ownership network $G_\alpha$.\par
The withholding tax rates imposed on dividends that a firm pays to a firm in other jurisdiction are defined in the domestic laws of each jurisdiction or the tax treaties concluded between jurisdictions. We used the reduced withholding tax rates because the purpose of our analysis is treaty shopping. The data has been extracted from Ernst \& Young (2017), which summarizes such withholding tax rates, for the withholding tax network $G_\beta$.

\section{Results and discussion}
At first, 25 jurisdiction $\times$ sector pairs are identified as "sinks." Next, the standardized conduit centrality $C_{js}$ clarified which jurisdictions $\times$ sector pairs through which more value passes and the standardized load centrality $L_j$ reveals which jurisdiction is attractive for treaty shopping. Finally, we calculated the standardized multilayer centrality $M_{js}$ by combining the standardized conduit centrality $C_{js}$ and the standardized load centrality $L_j$ and found that firms in certain sectors are often used for treaty shopping. Based on the result, we suggest countermeasures against treaty shopping to focus on certain sectors.

\subsection{Value intensity}
\subsubsection{Sink centrality}
\begin{figure}[ht]
\centering
\includegraphics[width=7cm]{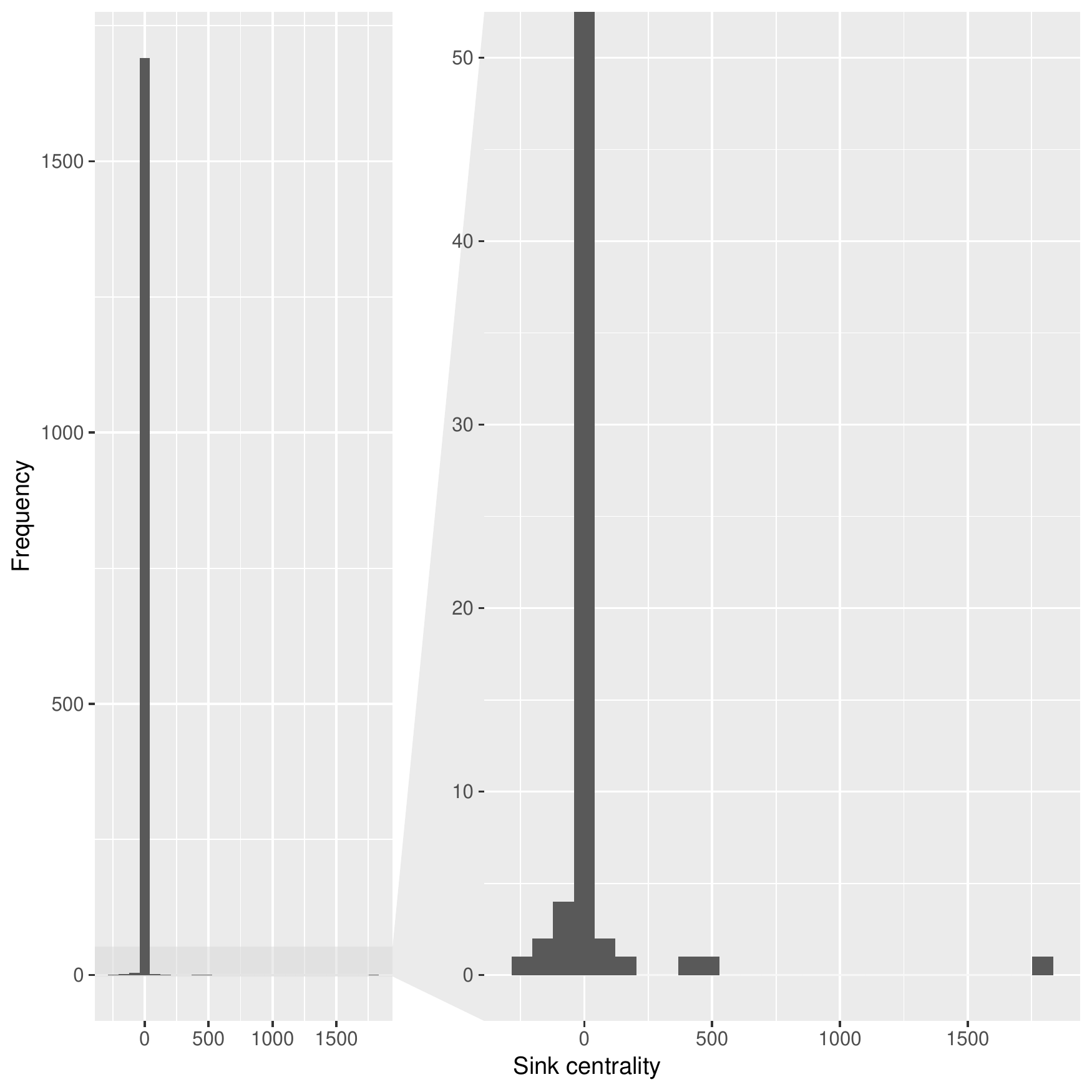}
\caption{{\bf Frequency distribution of sink centrality $S_{js}$.} The sink centrality of most jurisdictions $\times$ sector pairs is less than 1.0. The width of the bins is 81.5.}
\label{fig:4}       
\end{figure}

We calculated standardized sink centrality $S_{js}$ for 1,704 jurisdiction $\times$ sector pairs for which data can be obtained. The maximum value of the sink centrality was 1784.38 and the minimum value of that was -251.82. The value of sink centrality is wide, but most of the value of the sink centrality is concentrated in parts. Figure \ref{fig:4} shows its frequency distribution. The sink centrality of 1,598 pairs (about 93.8\% of the total pairs) is 1 or less. We can suppose that the pairs whose standardized sink centrality $S_{js}$ is high are very unique and are likely to function as a sink.

Table \ref{tab:1} shows 25 jurisdiction $\times$ sector pairs whose standardized sink centralities $S_{js}$ are higher than 10.0. Nine sectors are represented among the pairs with the highest sink centrality. Financial and insurance activity ("Finance \& Insurance") represents more than half of this list and professional, scientific, and technical activities ("Professional activities etc.") represents about 1/8 of this list, showing that much value remains in these sectors. For the calculation of standardized sink centrality $S_{js}$, we regarded the 25 pairs accounting for about 1.5\% of all pairs as sinks.

\begin{table}[ht]
\caption{Jurisdiction $\times$ Sector Pairs Where Sink Centrality $S_{js}$ is above 10}
\label{tab:1}       
\begin{tabular}{lll}
\hline\noalign{\smallskip}
Jurisdiction & Sector & $S_{js}$ \\
\noalign{\smallskip}\hline\noalign{\smallskip}
Malta & Finance \& Insurance & 1784.38 \\
Luxembourg & Professional activities etc. & 511.86 \\
Luxembourg & Administrative \& Support service & 369.72 \\
Bermuda & Construction & 134.63 \\
Bermuda & Finance \& Insurance & 87.75 \\
British Virgin Islands & Manufacturing & 57.03 \\
Cayman Islands & Finance \& Insurance & 39.75 \\
Curacao & Finance \& Insurance & 34.47 \\
France & Finance \& Insurance & 34.10 \\
Marshall Islands & Transportation \& Storage & 30.36 \\
Sweden & Finance \& Insurance & 28.00 \\
British Virgin Islands & Wholesale \& Retail trade & 23.13 \\
Cyprus & Finance \& Insurance & 19.56 \\
Spain & Finance \& Insurance & 19.04 \\
Curacao & Wholesale \& Retail trade & 16.43 \\
UK & Mining \& Quarrying & 16.18 \\
Portugal & Finance \& Insurance & 15.50 \\
Norway & Finance \& Insurance & 14.30 \\
Belgium & Finance \& Insurance & 13.07 \\
the UK & Finance \& Insurance & 12.95 \\
Austria & Professional activities etc. & 12.71 \\
Iceland & Finance \& Insurance & 12.69 \\
South Africa & Manufacturing & 10.81 \\
Singapore & Other service etc. &10.48 \\
UK & Professional activities etc. & 10.25 \\
\noalign{\smallskip}\hline
\end{tabular}
\end{table}

\subsubsection{Conduit outward and inward centrality}
\label{sec:4-1-2}
\begin{figure}[ht]
\centering
\includegraphics[width=7cm]{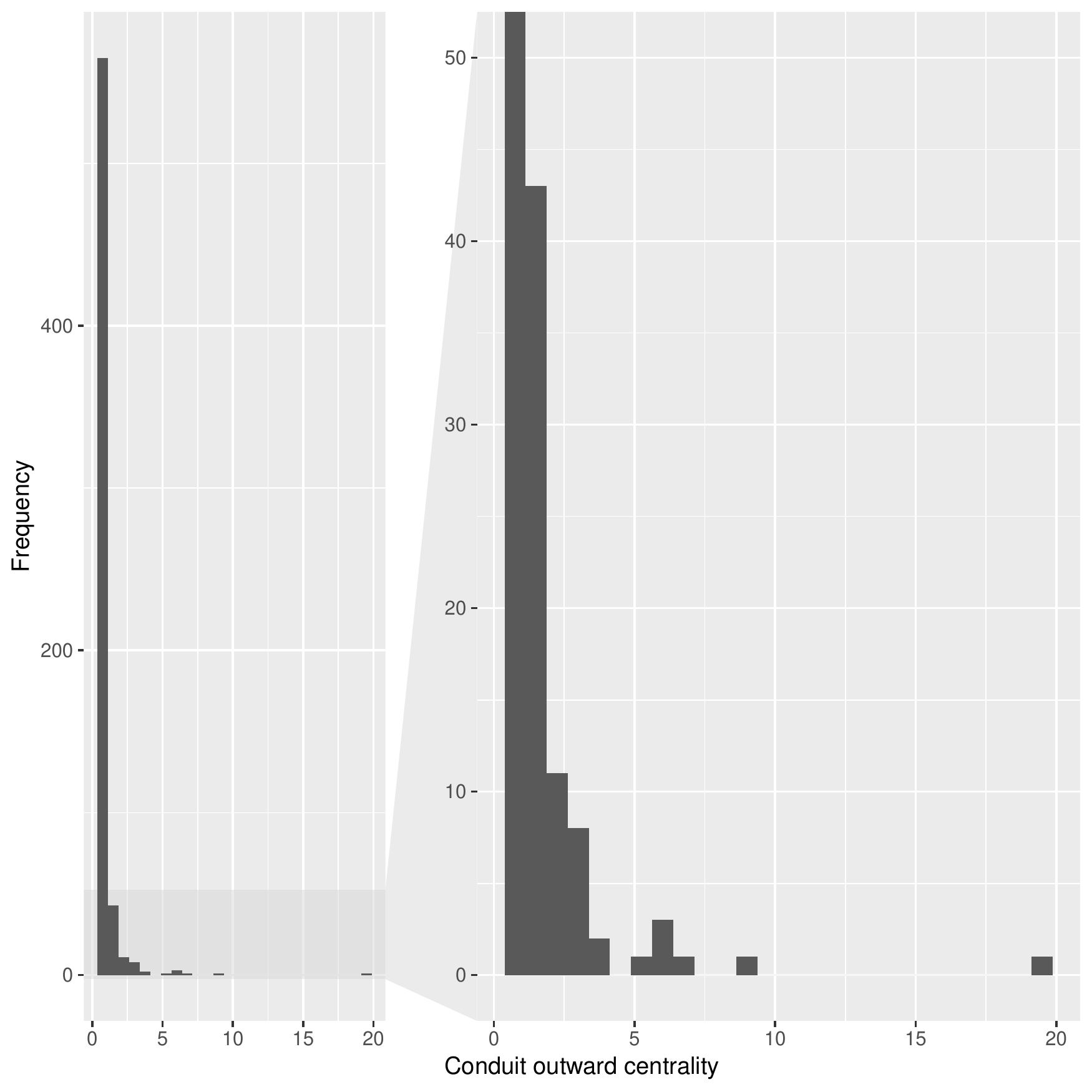}
\caption{{\bf Frequency distribution of the standardized conduit outward centrality $C^{out}_{js}$.} Most of the centralities are less than 1.0. The width of the bins is 1.1.}
\label{fig:5}       
\end{figure}

\begin{figure}[ht]
\centering
\includegraphics[width=7cm]{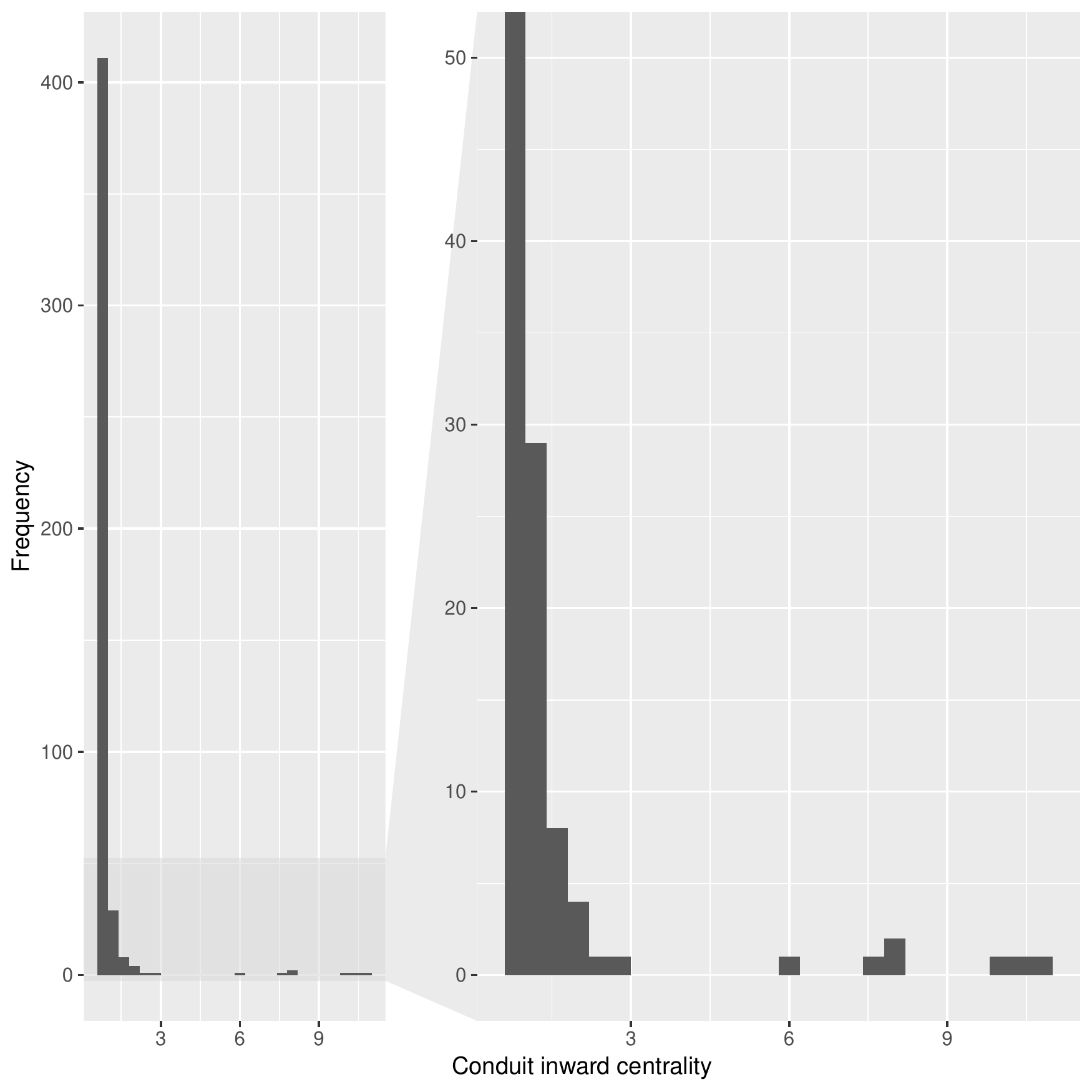}
\caption{{\bf Frequency distribution of the standardized conduit inward centrality $C^{in}_{js}$.} Most of the centralities are less than 1.0. The width of the bins is 1.1.}
\label{fig:6}       
\end{figure}

The standardized conduit outward centrality $C_{js}^{in}$ was calculated for 636 jurisdiction $\times$ sector pairs and the standardized conduit inward centrality $C_{js}^{in}$ was calculated for 461 jurisdiction $\times$ sector pairs. Finally, we obtained 389 pairs having both standardized conduit outward centralities $C_{js}^{out}$ and standardized conduit inward centralities $C_{js}^{in}$ . The maximum value of the standardized conduit outward centrality $C_{js}^{out}$ was 19.37 and the minimum value was 0.78. The maximum value of the standardized conduit inward centrality $C_{js}^{in}$ is 10.79 and the minimum value is 0.81. Even though the differences between their maximum value and their minimum value are not small, both centralities were distributed in certain parts. Figure \ref{fig:5} and Figure \ref{fig:6} show frequency distributions of the standardized conduit outward centrality $C^{out}_{js}$ and the standardized conduit inward centrality $C^{in}_{js}$ respectively. 608 pairs, accounting for 95.6\% of the total pairs, have standardized conduits outward centrality $C_{js}^{out}$ less than 1.0. Similarly, 449 pairs, accounting for 97.4\% of the total pairs, have standardized conduit inward centrality $C_{js}^{in}$ less than 1.0. We can suppose that the pairs whose standardized conduit centrality $C_{js}$ is high are very unique and are likely to function as conduits.

Table \ref{tab:2} shows 15 jurisdiction $\times$ sector pairs whose standardized conduit centrality $C_{js}$ is over 2.0. Seven sectors across 13 jurisdictions are represented among the 15 pairs. Most of the 13 jurisdictions are developed countries and there are no small island jurisdictions except Bermuda. Wholesale and retail trade and repair of motor vehicles and motorcycles ("Wholesale \& Retail trade etc.") represents more than 1/4 of this list and Finance \& Insurance represents about 1/5 of this list, showing that considerable value, either from or toward sinks, passes through firms in these sectors.

\begin{table}
\caption{Jurisdiction $\times$ Sector Pairs Whose Standardized Conduit Centrality $C_{js}$ is over 2.0}
\label{tab:2}       
\begin{tabular}{llllll}
\hline\noalign{\smallskip}
Jurisdiction & Sector & $C^{out}_{js}$ & $C^{in}_{js}$ & $C_{js}$ & \\
\noalign{\smallskip}\hline\noalign{\smallskip}
The Netherlands & Finance \& Insurance & 19.37 & 6.13 & 14.36 \\
Luxembourg & Wholesale and Retail trade etc. & 6.18 & 10.79 & 8.79 \\
Bermuda & Mining \& Quarrying & 1.33 & 9.49 & 7.48 \\
Luxembourg & Finance \& Insurance & 6.55 & 8.08 & 7.36 \\
Sweden & Electricity \& Gas supply etc. & 9.03 & 0.91 & 6.41 \\
Austria & Wholesale \& Retail trade etc. & 3.21 & 7.87 & 6.01 \\
Bermuda & Wholesale \& Retail trade etc. & 4.12 & 7.42& 6.00 \\
Portugal & Professional activities etc. & 6.35 & 1.24 & 4.57 \\
Malaysia & Manufacturing & 5.62 & 1.03 & 4.04 \\
Switzerland & Wholesale \& Retail trade etc. & 3.85 & 0.93 & 2.80 \\
Germany	& Manufacturing & 3.08 & 2.19 & 2.46 \\
UK & Administrative \& Support services & 1.86 & 2.94 & 2.46 \\ 
France & Professional activities etc. & 3.23 & 1.05 & 2.40 \\
Ireland & Finance \& Insurance & 3.12 & 1.18 & 2.36 \\
Austria & Mining \& Quarrying & 3.08 & 0.84 & 2.26 \\
Austria & Manufacturing & 2.53 & 1.61 & 2.12 \\
Portugal & Information \& Communication & 2.16 & 1.94 & 2.05 \\
\noalign{\smallskip}\hline
\end{tabular}
\end{table}

\subsection{Withholding tax rate}
\begin{figure}[ht]
\centering
\includegraphics[width=7cm, angle=0]{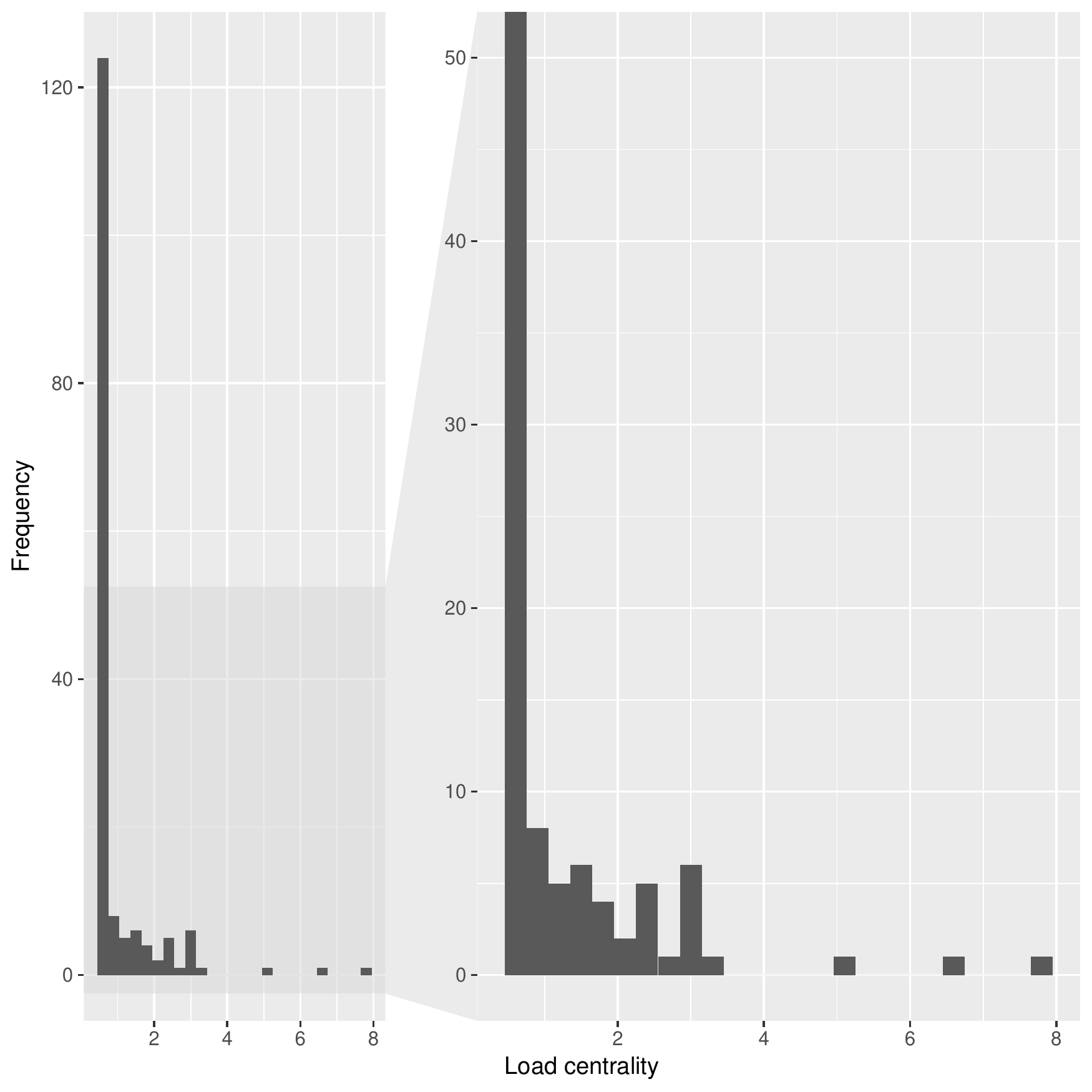}
\caption{{\bf Distribution of standardized load centrality $L_j$.} Most of the centralities are below 1.0.}
\label{fig:7}       
\end{figure}

We calculated the standardized load centrality $L_j$ for 165 jurisdictions to find which jurisdictions are likely to be used for treaty shopping. The standardized load centrality $L_j$ considers all withholding tax liabilities imposed on dividends made between 165 jurisdictions (27,060 pairs in total). The maximum value of the standardized load centrality $L_j$ was 7.87 and the minimum value was 0.59. The value of the standardized load centrality $L_j$ is wide, but most of the value of the standardized load centrality $L_j$ is concentrated in parts. Figure \ref{fig:7} shows its relative frequency distribution. The standardized load centrality $L_j$ of 147 jurisdictions (about 89.0\% of 165 jurisdictions) is 1 or less and the jurisdictions with high standardized load centralities $L_j$ are limited. We can suppose that the jurisdictions whose standardized load centrality $L_j$ is high are likely to be used for treaty shopping.

Table \ref{tab:3} shows jurisdictions whose standardized load centrality $L_j$ is in the top 15. It is confirmed that focusing on withholding tax rates imposed on dividends and evaluating jurisdictions by the load centralities $l_j$ is meaningful to the analysis of treaty shopping because the list contains jurisdictions that multinationals usually use for treaty shopping (Diamond et al. 2017; Nakamoto and Ikeda 2018). The 15 jurisdictions are scattered around Europe, the Middle East, East Asia, Africa, and America. European jurisdictions especially have high standardized load centrality $L_j$ because the withholding tax imposed on dividends made between EU member states is exempted by the European Union directive (Directive 90/435/EC).

\begin{table}[ht]
\caption{Jurisdictions Whose Standardized Load Centralities $L_j$ are in the Top 15}
\label{tab:3}       
\begin{tabular}[h]{llllll}
\hline\noalign{\smallskip}
Jurisdiction & $L_j$ & Jurisdiction & $L_j$ & Jurisdiction & $L_j$ \\
\noalign{\smallskip}\hline\noalign{\smallskip}
UK & 7.87 & Singapore & 2.94 & Saint Lucia & 2.46 \\
UAE & 6.59 & Switzerland & 2.93 & Bahrain & 2.46 \\
Kuwait & 5.22 & Mauritius & 2.91 & Malaysia & 2.45 \\
The Netherlands & 3.44 & Spain & 2.86 & Ireland & 2.39 \\
Cyprus & 3.11 & Luxembourg & 2.65 & Estonia & 2.24 \\
Hong Kong & 3.05 & Qatar & 2.44 & Malta & 2.21 \\
\noalign{\smallskip}\hline
\end{tabular}
\end{table}

\subsection{Multilayer centrality}
The standardized multilayer centrality $M_{js}$ can be calculated regarding 389 pairs because the calculation needs both the standardized conduit outward centrality $C^{out}_{js}$ and the standardized conduit inward centrality $C^{in}_{js}$. In addition, for the calculation of the standardized multilayer centrality $M_{js}$, $\alpha$ and $\beta$ need to be set as in Eqs. (\ref{eq:8}) and (\ref{eq:7}). $\alpha$ and $\beta$ determine how much value intensity and withholding tax rate are considered, respectively. We tried to find the appropriate value of $\beta$ by comparing the results obtained by fixing $\alpha$ to 1.0 and fluctuating $\beta$. Table \ref{tab:4} contains the jurisdiction $\times$ sector pairs whose standardized multilayer centrality $M_{js}$ is above 2.0 when $\beta$ is set to 0.8. Many listed pairs are in the United Kingdom and we cannot see the difference between sectors because the standardized load centrality $L_j$ does not take the difference of sectors into account. In other words, it shows that setting $\beta = 0.8$ takes too much of the withholding tax rate into consideration. Table \ref{tab:5} shows the pairs whose standardized multilayer centrality $M_{js}$ is above 2.0 when $\beta$ is set to 0.1. The listed pairs include jurisdictions such as Germany and France, which are unlikely to be used for treaty shopping because these jurisdictions are not listed in Table \ref{tab:2}. There may be another reason (except treaty shopping) why more value passes through such jurisdictions.

\begin{table}
\caption{Jurisdiction $\times$ Sector Pairs Whose Standardized Multilayer Centrality $M_{js}$ is above 2.0 ($\beta = 0.8$)}
\label{tab:4}       
\begin{tabular}[h]{llllll}
\hline\noalign{\smallskip}
Jurisdiction & Sector & $M^{out}_{js}$ & $M^{in}_{js}$ & $M_{js}$ \\
\noalign{\smallskip}\hline\noalign{\smallskip}
The Netherlands & Finance \& Insurance & 8.98 & 4.74 & 7.18\\
Luxembourg & Wholesale \& Retail trade etc. & 4.24 & 5.78 & 5.07\\
Luxembourg & Finance \& Insurance & 4.38 & 4.93 & 4.66\\
UK & Administrative activities etc. & 3.53 & 4.55 & 4.08\\
UK & Manufacturing & 3.13 & 3.18 & 3.15\\
Malaysia & Manufacturing & 3.89 & 1.52 & 2.95\\
Switzerland & Wholesale \& Retail trade etc. & 3.41 & 1.55 & 2.65\\
UK & Information \& Communication & 2.40 & 2.44 & 2.42\\
UK & Electricity \& Gas supply etc. & 2.37 & 2.41 & 2.39\\
Bermuda & Mining \& Quarrying & 1.01 & 3.19 & 2.37\\
UK & Wholesale \& Retail trade etc.	& 2.29 & 2.42 & 2.35\\
Sweden & Electricity \& Gas supply etc.	& 3.18 & 0.89 & 2.33\\
Bermuda & Wholesale \& Retail trade etc. & 1.90	& 2.63 & 2.29\\
UK & Construction & 2.22 & 2.35 & 2.29\\
UK & Transportation \& Storage & 2.23 & 2.32 & 2.27\\
Ireland & Finance \& Insurance & 2.77 & 1.62 & 2.27\\
UK & Water supply etc. & 2.20 & 2.30 & 2.25\\
UK & Other service & 2.20 & 2.30 & 2.25\\
UK & Real estate activities	& 2.23 & 2.25 & 2.24\\
UK & Arts \& Entertainment etc. & 2.20 & 2.28 & 2.24\\
UK & Public Administration etc. & 2.24 & 2.23 & 2.24\\
Singapore & Wholesale \& Retail trade etc. & 2.41 & 2.05 & 2.23\\
UK & Human health etc. & 2.18 & 2.24 & 2.21\\
UK & Accommodation and Food service etc. & 2.19 & 2.24 & 2.21\\
UK & Education & 2.18 & 2.24 & 2.21\\
UK & Agriculture etc. & 2.18 & 2.23	& 2.21\\
Austria & Wholesale \& Retail trade etc. & 1.60 & 2.64 & 2.18\\
Singapore & Manufacturing & 1.87 & 2.44 & 2.17\\
Ireland & Professional activities etc. & 2.40 & 1.80 & 2.12\\
Spain & Professional activities etc. & 2.07 & 1.94 & 2.00\\
\noalign{\smallskip}\hline
\end{tabular}
\end{table}

\begin{table}
\caption{Jurisdiction $\times$ Sector Pairs Whose Standardized Multilayer Centrality $M_{js}$ is above 2.0 ($\beta = 0.1$)}
\label{tab:5}       
\begin{tabular}[ht]{llllll}
\hline\noalign{\smallskip}
Jurisdiction & Sector & $M^{out}_{js}$ & $M^{in}_{js}$ & $M_{js}$ & \\
\noalign{\smallskip}\hline\noalign{\smallskip}
The Netherlands& Finance \& Insurance & 13.00 & 5.37 & 9.94 \\
Luxembourg & Wholesale \& Retail trade etc. & 5.08 & 7.81 & 6.59 \\
Luxembourg & Finance \& Insurance & 5.32 & 6.25 & 5.80 \\
Bermuda & Mining \& Quarrying & 1.16 & 5.65 & 4.08 \\
Sweden & Electricity \& Gas supply etc. & 5.25 & 0.90 & 3.77 \\
Bermuda & Wholesale \& Retail trade etc. & 2.75 & 4.33 & 3.63 \\
Austria & Wholesale \& Retail trade etc. & 2.24 & 4.46 & 2.20 \\
Malaysia & Manufacturing & 4.64 & 1.26 & 3.40 \\
UK & Administrative \& Support service & 2.60 & 3.69 & 3.19 \\
Switzerland & Wholesale \& Retail trade etc. & 3.62 & 1.21 & 2.70 \\
Portugal & Professional activities etc. & 3.66 & 1.04 & 2.69 \\
Ireland & Financial \& Insurance & 2.93 & 1.39 & 2.29 \\
UK & Manufacturing & 2.20 & 2.24 & 2.22 \\
Germany & Manufacturing & 2.43 & 1.87 & 2.17 \\
France & Professional activities etc. & 2.72 & 1.14 & 2.08 \\
Ireland & Professional activities etc. & 2.41 & 1.61 & 2.05 \\
Singapore & Wholesale \& Retail trade etc. & 2.23 & 1.78 & 2.02 \\
\noalign{\smallskip}\hline
\end{tabular}
\end{table}

Table \ref{tab:6} shows pairs whose standardized multilayer centrality $M_{js}$ is above 2.0 when $\beta$ is set to 0.5 and includes the jurisdictions well-known as conduits (Diamond et al. 2017) and, therefore, likely to be used for treaty shopping. When we set $\beta$ to 0.3 for our analysis of conduit firms, 179 pairs had standardized multilayer centrality $M_{js}$ above 1.0, accounting for 46.02\% of total pairs; 48 pairs had standardized multilayer centrality $M_{js}$ above 1.5, accounting for 12.34\%; and 16 pairs had standardized multilayer centrality $M_{js}$ above 2.0, accounting for 4.11\%. Finally, we decided to consider the pairs whose standardized multilayer centrality $M_{js}$ is above 2.0 as the pairs that are likely to be used for treaty shopping.\par
The 13 pairs listed in Table \ref{tab:6} include eight sectors across 11 jurisdictions and Finance \& Insurance and Wholesale \& Retail trade etc., firms are remarkable in Table \ref{tab:6} compared with other sectors. This implies that these sectors have greater potential to be used for treaty shopping. It is possible that treaty shopping can be prevented by making new tax rules focusing on these sectors. In addition, no Middle East Asian jurisdictions are ranked highly in the standardized multilayer centrality $M_{js}$, although their standardized load centralities $L_{j}$ are ranked highly (see Table \ref{tab:3}). Other regulations of the jurisdictions imposed may cause the results.\par

\begin{table}[ht]
\caption{Jurisdiction $\times$ Sector Pairs Whose Standardized Multilayer Centrality $M_{js}$ is above 2.0 ($\beta = 0.5$)}
\label{tab:6}       
\begin{tabular}[h]{llllll}
\hline\noalign{\smallskip}
Jurisdiction & Sector & $M^{out}_{js}$ & $M^{in}_{js}$ & $M_{js}$ \\
\noalign{\smallskip}\hline\noalign{\smallskip}
The Netherlands & Finance \& Insurance & 10.88 & 5.06 & 8.49 \\
Luxembourg & Wholesale \& Retail trade & 4.66 & 6.76 & 5.81 \\
Luxembourg & Finance \& Insurance & 4.85 & 5.57 & 5.22 \\
UK & Administrative \& Support services & 3.01 & 4.08 & 3.59 \\
Malaysia & Manufacturing & 4.26 & 1.38 & 3.17 \\
Bermuda & Mining \& Quarrying & 1.09 & 4.30 & 3.13 \\
Sweden & Electricity \& Gas supply etc. & 4.13 & 0.89 & 2.99 \\
Bermuda & Wholesale \& Retail trade & 2.30 & 3.41 & 2.91 \\
Austria & Wholesale \& Retail trade & 1.90 & 3.46 & 2.80 \\
Switzerland & Wholesale \& Retail trade & 3.52 & 1.36 & 2.67 \\
UK & Manufacturing & 2.60 & 2.65 & 2.63 \\
Ireland & Finance \& Insurance & 2.85 & 1.49 & 2.28 \\
Singapore & Manufacturing & 1.71 & 2.35 & 2.05 \\
\noalign{\smallskip}\hline
\end{tabular}
\end{table}

At present, the mainstream countermeasures against treaty shopping involve introducing a limitation of benefit clause or a principal purpose test into tax treaties. The limitation of benefit clause limits firms that can receive the withholding tax reduction or exemption by certain criteria. On the other hand, the principal purpose test deprives firms whose main purpose is to enjoy the reduction or exemption of withholding tax liabilities. Developing countries prefer the principal purpose test because they are easy to enforce compared with the limitation of benefit clause, whose application criteria are complicated. On the other hand, the business community is concerned with the principal purpose test because the test is unclear as to the main purpose and tends to prefer the limitation of benefit clause whose application criteria are clearer.\par
Our analysis shows that some sectors, such as Manufacturing, Wholesale \& Retail trade etc., Professional activities etc., are likely to be used for treaty shopping. We think that it is effective for preventing treaty shopping to focus on such sectors as Controlled Foreign Company rules of some jurisdictions, which is a countermeasure against another scheme of international tax avoidance, already focused on sectors. The introduction of rules focusing on some sectors may not only prevent treaty shopping effectively but also reduce the complexity of application criteria developing countries are concerned with and improve taxpayer predictability.\par
\begin{figure}[ht]
\centering
\subfigure[Finance \& Insurance]{
\includegraphics[width=8cm]{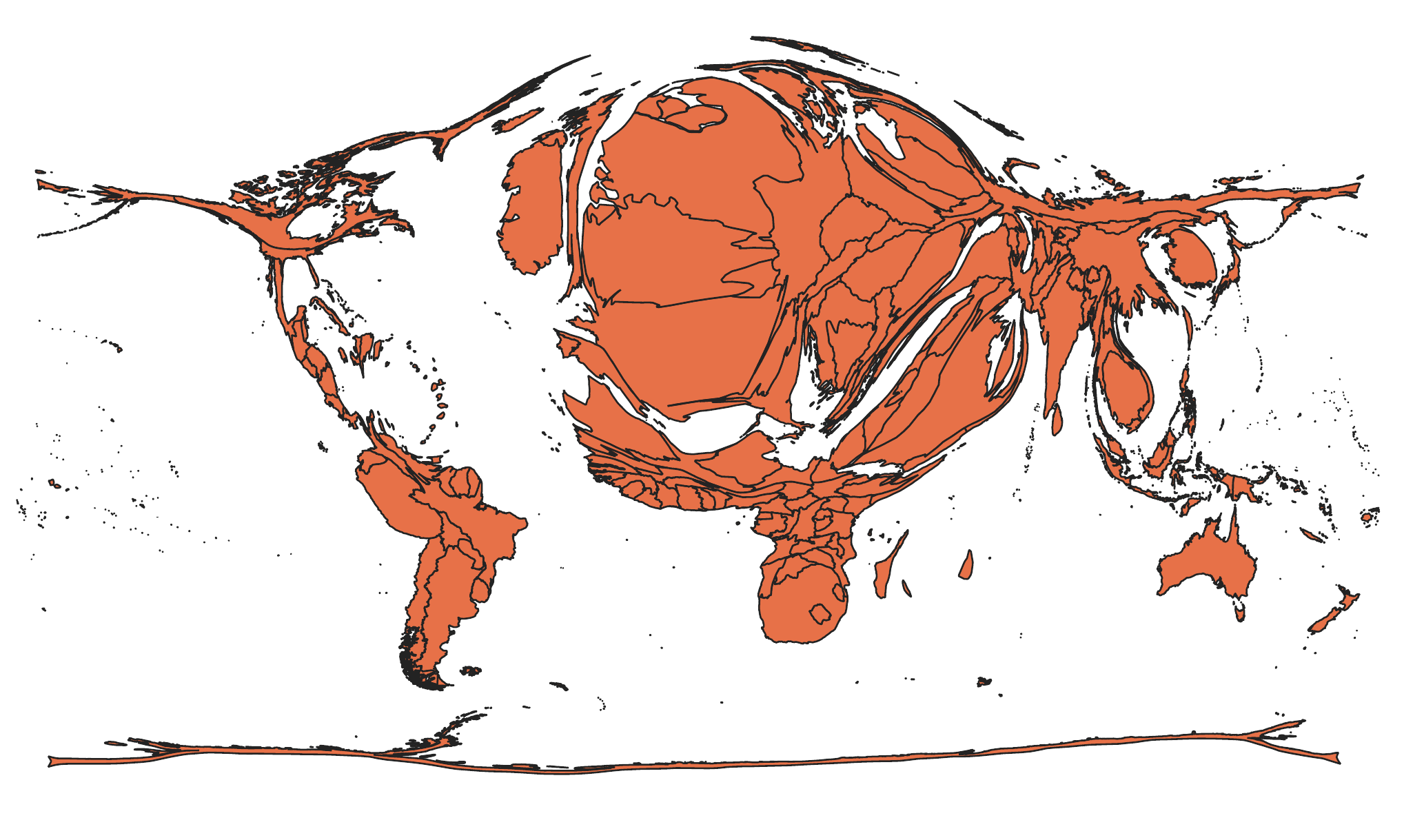}
}
\subfigure[Wholesale \& Retail trade etc.]{
\includegraphics[width=8cm]{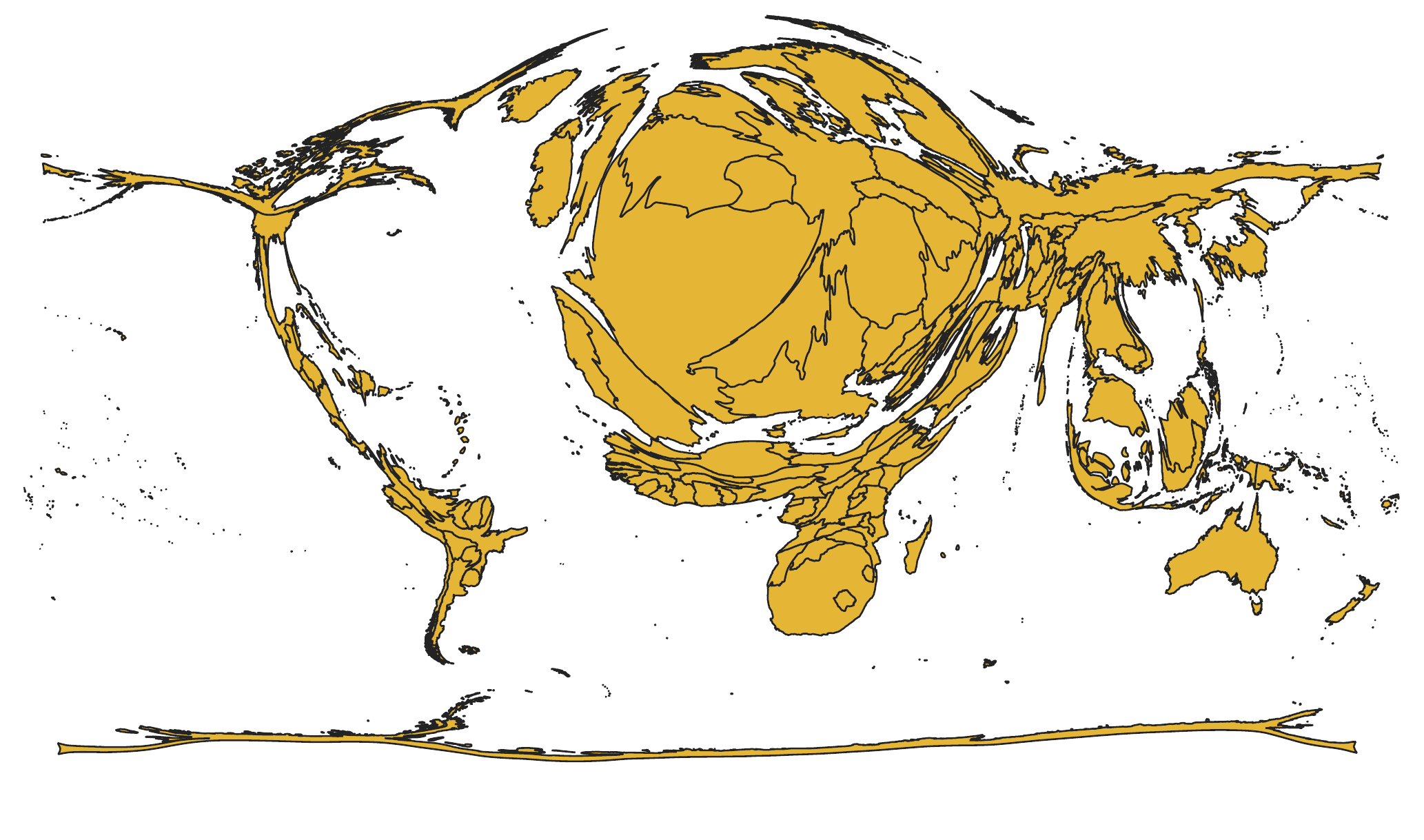}
}
\caption{{\bf Cartogram of multilayer centralities $M_{js}$.} The size of each jurisdiction indicates the degree of multilayer centrality $M_{js}$.}
\label{fig:8}       
\end{figure}

The size of each jurisdictions in Figure \ref{fig:8} indicates the size of the standardized multilayer centrality $M_{js}$ regarding Finance \& Insurance and Wholesale \& Retail trade etc., respectively. It should be noted that jurisdictions with high standardized multilayer centrality $M_{js}$ are limited. Figure \ref{fig:8} (a) indicates the standardized multilayer centrality $M_{js}$ of Finance \& Insurance and shows that the centrality of The Netherlands, Luxembourg, and other financial centers such as the United Kingdom, Bahrain, Hong Kong, and Mauritius is high. Figure \ref{fig:8} (b) indicates the standardized multilayer centrality $M_{js}$ of Wholesale \& retail trade etc., and shows that the centrality of the jurisdictions of Europe and South East Asia is high. We suggest that the new clauses focusing on certain sectors are introduced to tax treaties already concluded with jurisdictions having high standardized multilayer centrality $M_{js}$.  

Developing countries conclude tax treaties with developed countries to increase investment from developed countries. With globalization of economy, the number of tax treaties has increased and about 1,000 tax treaties are related to developing countries among about 4,000 tax treaties in the world. It should be noted that tax treaties not only increase foreign direct investment from developed countries, but also increase the possibility of treaty shopping and the loss of their own tax sources.


\section{Conclusion}
Developing countries have mainly relied on assistance from developed countries to reduce poverty, but they are now required to carry out their own development through their own tax revenues. International tax avoidance is one of the reasons why developing countries cannot earn tax revenues compared with developed countries. Even though there are various ways to conduct international tax avoidance, treaty shopping is focused on in this paper due to tax revenues of developing countries. To analyze the location and sector of conduit firms that are likely to be used for treaty shopping, we constructed the multilayer ownership-tax network and proposed the multilayer centrality. Because multilayer centrality can consider not only the value flowing in the ownership network but also the withholding tax rate, it is expected to grasp precisely the locations and the sectors of conduit firms established for the purpose of treaty shopping. The results of our analysis suggest that firms in the sectors of Finance \& Insurance and Wholesale \& Retail trade etc. may be conduit firms that plays an important role in treaty shopping. Therefore, we suggest that to prevent treaty shopping, developing countries should introduce a clause focusing on certain sectors in their tax treaty, especially with developed countries with high multilayer centrality. This is because the countermeasures to treaty shopping that focus on some sectors is not complicated, developing countries find it easier to use them, and the predictability of taxpayers is not harmed much. A further quantitative study of treaty shopping is needed that takes not only withholding tax but also corporate tax into consideration. Such findings would contribute to our understanding of the effects of treaty shopping toward each jurisdiction's tax revenue.

\begin{acknowledgements}
I would like to thank Tadao Okamura and Hiroaki Takashima for valuable comments.
\end{acknowledgements}
{\footnotesize {\bf Conflict of interest statement} On behalf of all authors, the corresponding author states that there is no conflict of interest.}

\end{document}